\documentclass[runningheads]{llncs}

\usepackage[english]{babel}

\usepackage{authblk}

\usepackage{graphicx}
\graphicspath{ {img/} }

\usepackage[colorlinks=true, allcolors=black]{hyperref}

\usepackage{tikz}
\usetikzlibrary{quantikz2}

\usepackage{xcolor}
\usepackage[normalem]{ulem}
\usepackage{subcaption}

\usepackage[T1]{fontenc}

\begin{document}

\title{Evaluating Quantum Wire Cutting for QAOA: Performance Benchmarks in Ideal and Noisy Environments}
\titlerunning{Evaluating Quantum Wire Cutting for QAOA}

\author{R. M. J. (Michel) Meulen\inst{1,2} \and Niels M. P. Neumann\inst{1} \and Jasper Verbree\inst{1}} 
\institute{Department of Applied Cryptograhpy \& Quantum Algorithms, The Netherlands Organisation for Applied Scientific Research, The Netherlands \email{\{\textit{first\_name}.\textit{last\_name}\}@tno.nl}\and
Fontys University of Applied Sciences, Eindhoven, The Netherlands}
\authorrunning{R. M. J. Meulen, N. M. P. Neumann and J. Verbree}

\maketitle

\begin{abstract}
    Current quantum computers suffer from a limited number of qubits and high error rates, limiting practical applicability.
    Different techniques exist to mitigate these effects and run larger algorithms. 
    In this work, we analyze one of these techniques called quantum circuit cutting. 
    With circuit cutting, a quantum circuit is decomposed into smaller sub-circuits, each of which can be run on smaller quantum hardware. 
    We compare the performance of quantum circuit cutting with different cutting strategies, and then apply circuit cutting to a QAOA algorithm. 
    Using simulations, we first show that Randomized Clifford measurements outperform both Pauli and random unitary measurements.
    Second, we show that circuit cutting has trouble providing correct answers in noisy settings, especially as the number of circuits increases. 
\keywords{Quantum Computing \and Circuit Cutting \and Wire Cutting \and QAOA \and MaxCut.}
\end{abstract}

\section{Introduction}\label{sec:Introduction}
Quantum computers have the potential to solve problems infeasible for any classical supercomputer~\cite{McArdle:2020,Shor:1995}. 
Investments in quantum computers and the corresponding algorithms are continuously rising. 
However, current quantum computers, often called Noisy Intermediate-Scale Quantum (NISQ) devices, are still far away from their full potential. 
They have a limited number of qubits and high error rates~\cite{preskill2018quantum}.
Improving usability requires increasing the number of qubits and decreasing the error rates.
Recently, techniques like error mitigation have been proposed to mitigate, rather than solve, these limitations~\cite{Kandala:2019,TemmeBravyiGambetta:2017}.

In 2020, Peng et al. showed how to use an $n$-qubit quantum computer to run an $n+k$-qubit quantum circuit~\cite{PengTimeWiseCut}, introducing the subject of \emph{circuit cutting}.
Their method decomposes a quantum circuit in different blocks and combines the outcomes of the individual blocks to approximate expectation values produced by the original circuit.
Follow-up work improved their method and provided explicit decompositions for certain quantum gates~\cite{Bechtold2023_CuttingPatterns,RandomizedMeasurementsWireCutting,Mitarai:2021}. 
Typically, we distinguish between two different circuit cutting techniques: wire cutting, where a qubit is measured in a certain basis and initialized again, and gate cutting, where a multi-qubit quantum gate is replaced by single-qubit gates and measurements with post-selection~\cite{Bechtold2023_CuttingPatterns}. 
This work focuses on wire cutting. 

First, we compare three variants of wire cutting, using exact and noisy simulations, and determine which of these variants performs best. 
Second, we apply the best wire cutting variant found to the quantum approximate optimization algorithm (QAOA)~\cite{farhi2014quantumapproximateoptimizationalgorithm} applied to the MaxCut problem~\cite{PhysRevX.10.021067}.
In this way, we learn the potential of quantum circuit cutting and its overhead for a practical problem
showing how the exact theory translates to the noisy reality.

The remainder of this paper is organized as follows:
Section \ref{sec:background} discusses in more detail wire cutting and the variants considered in this work.
Section \ref{sec:methodology} presents the experimental setup: a benchmark comparison between Pauli-based and randomized measurement-based wire cutting; and the application of the randomized measurement-based wire cutting technique to QAOA in both ideal and noisy simulation environments. 
Section \ref{sec:results} gives the results for these experiments, and Section \ref{sec:conclusion} discusses the results and concludes the work.

\section{Wire cutting}\label{sec:background}
Tang et al. introduced a new method for wire cutting based on decomposing quantum operations into linear combinations of Pauli matrices~\cite{CutQC}. 
By measuring qubits in the $X$, $Y$, and $Z$ bases and reinitializing them in the $\ket{0}$, $\ket{1}$, $\ket{+}$, or $\ket{i}$ state, a larger circuit can be approximated by smaller ones. 

\begin{figure*}[t]
    \centering
    \includegraphics[width=0.8\linewidth]{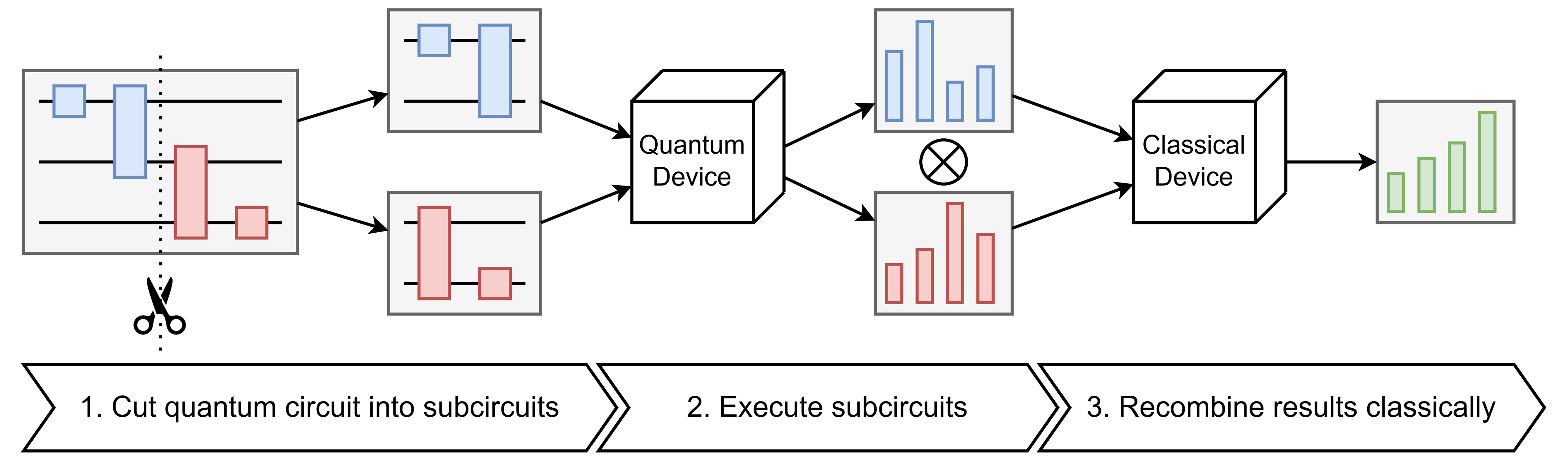}
    \caption{Visual representation of wire cutting using Pauli measurements. 
    Each subcircuit is run on a quantum device and the results are classically recombined.
    The tensor product symbol ($\otimes$) refers to the recombination of the different probability distributions as in Equation~\eqref{eq:pauli_wire_cut}.}
    \label{fig:pauli_measurements_wire_cutting} 
\end{figure*}

Once all measurement outcomes are obtained, the probability of a basis state being produced by the uncut circuit can be reconstructed. 
For instance, consider the state $\ket{010}$ where the second wire is cut. 
We define two auxiliary states $u_1 = \ket{00}$ and $u_2 = \ket{01}$ for the first and second qubit, and an output state $d = \ket{10}$ for the second and third qubit.
The probability of measuring $\ket{010}$ is:
\begin{equation}
\mathrm{Pr}[\ket{010}] = \frac{1}{2} \sum_{i=1}^{4}{U_i D_i},
\label{eq:pauli_wire_cut}
\end{equation}
where
\begin{align*}
    &U_1 = 2u_{1,Z}, \quad U_2 = 2u_{2,Z}, \quad U_3 = u_{1,X} - u_{2,X}, \quad U_4 = u_{1,Y} - u_{2,Y},  \\
     &D_1 = d_{\ket{0}}, \quad D_2 = d_{\ket{1}}, \quad D_3 = 2d_{\ket{+}} - d_{\ket{0}} - d_{\ket{1}}, \quad D_4 = 2d_{\ket{i}} - d_{\ket{0}} - d_{\ket{1}},
\end{align*}
and $u_{i,P}$ corresponds to the probability of measuring state $u_i$ in Pauli basis~$P$ and $d_{\ket{\phi}}$ corresponds to the probability of measuring state $d$ on input state $\ket{\phi}$.
A schematic presentation of wire cutting is given in \autoref{fig:pauli_measurements_wire_cutting}. Note that cutting the middle qubit wire of a three-qubit circuit results in subcircuits of two qubits.

\begin{figure*}[t]
    \centering
    \includegraphics[width=0.8\linewidth]{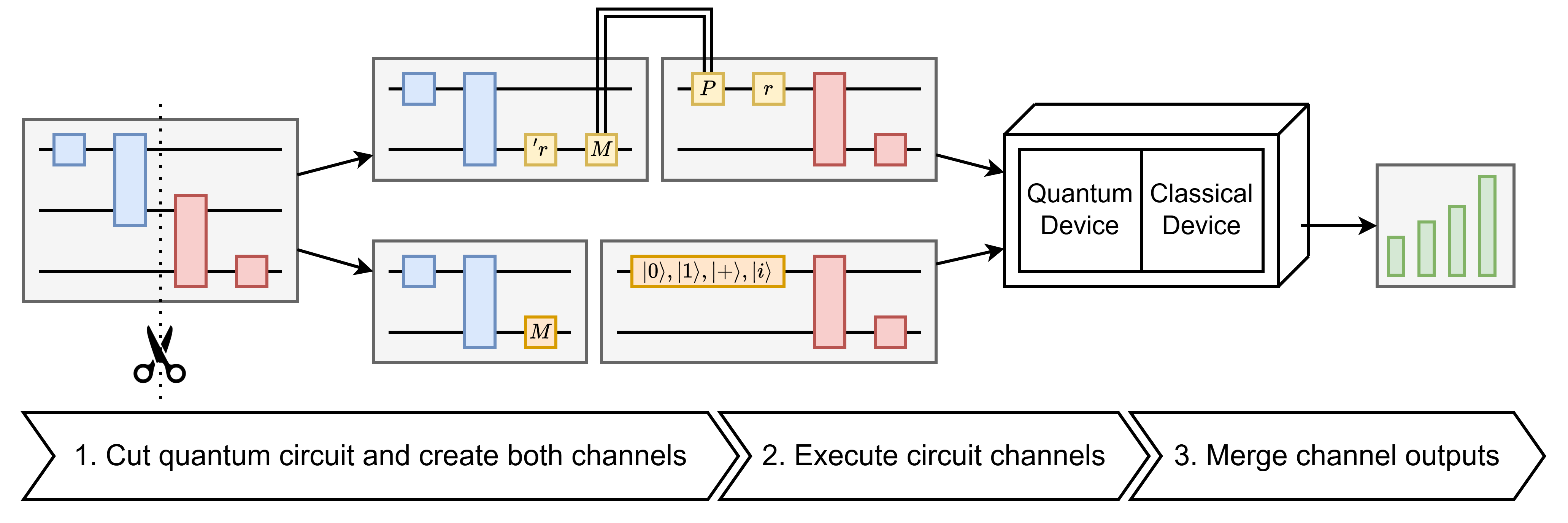}
    \caption{Visual representation of randomized-measurement-based wire cutting. 
    For every sample, the method randomly chooses between two quantum channels: 
    The random Clifford channel (top) and the depolarization channel (bottom).
    The chosen channel is then executed on a hybrid-quantum-classical-device.
    The recombination of the probabilities follows Equation~\eqref{eq:randomized_cutting_prob}.}
    \label{fig:random_measurements_wire_cutting}
\end{figure*}

Lowe et al. introduced an alternative approach using two distinct quantum channels to reduce the sampling overhead of exact methods~\cite{RandomizedMeasurementsWireCutting}. 
Their method is based on already existing methods to classically simulate the entanglement between matrix-product states~\cite{Chen:2021,Huang2020}.
One channel applies a random Clifford gate before the measurement and applies the conjugate of the Clifford gate as qubit reinitialization.
The other channel acts as a depolarizing channel, where the qubit is measured and then reinitialized in a random basis state. 
Their method randomly selects between these two channels with probabilities $\frac{d+1}{2d+1}$ and $\frac{d}{2d+1}$, where $d=2^k$ for $k$ cut qubits.
The probability corresponding to a state $b$ is then given by 
\begin{equation}
    \mathrm{Pr}(b) = (d+1)\cdot \mathrm{Pr}_{Clifford}(b) - d\cdot \mathrm{Pr}_{depolarizing}(b),
    \label{eq:randomized_cutting_prob}
\end{equation} 
where $\mathrm{Pr}_X(b)$ is the probability found for state $b$ for channel $X$. 
\autoref{fig:random_measurements_wire_cutting} gives an abstract overview of this randomized circuit cutting. 

The Clifford gates used in the random measurement circuit cutting approach are examples of unitary $2$-designs~\cite{RandomizedMeasurementsWireCutting}. 
One might wonder whether other gates improve the results even further. 
A simple generalization of Pauli gates is rotational Pauli gates. 
Even though these gates no longer form unitary $2$-designs, it is worth investigating whether they still help reduce the sampling overhead. 
We therefore also consider a modified random measurement circuit cutting approach where we apply random Pauli rotation gates before the measurement and initialize the qubits using its conjugated form. 

\section{Experimental setup}
\label{sec:methodology}
We now detail the experiments we have run and their implementation framework.

\subsection{Comparing circuit cutting methods}
In the first experiment we compared the exact wire cutting method based on Pauli measurements~\cite{CutQC} with the randomized measurement approach~\cite{RandomizedMeasurementsWireCutting} and a version where we apply arbitrary Pauli rotation gates. 
We consider a simple circuit that prepares a five-qubit GHZ state~\cite{greenberger1989going}, which we chose due to its wide applicability in many applications such as distributed quantum computing.
Additionally, the cascade-like nature of the quantum circuit makes it suitable for wire cutting. 
We added rotation gates before and after the CNOT gates to complicate the state and assure all measurement bases and initial states contribute. 

\subsection{Evaluating the performance for circuit cutting in QAOA}
In the second experiment, we use the best performing method to test the performance for a QAOA algorithm applied to the MaxCut problem.
Given a graph, the MaxCut problem is to partition the vertex set of the graph in two such that the number of edges with vertices in distinct partitions is maximized~\cite{MaxCutOriginJerzyAFilar2018NumericalAlgebraControlandOptimization}. 
For the QAOA algorithm, we assigned every vertex in the graph a qubit in the circuit. 
The cost operator is implemented by applying parametrized $ZZ$-gates between two qubits if and only if they are neighbors in the original graph. 
The mixing operator is implemented using parametrized $R_X$-gates. 
We implemented the parametrized $ZZ$-gates using $R_Z$-gates conjugated by CNOT-gates (where the $R_Z$-gate is applied to the target qubits of the CNOT-gates). 

\begin{figure}[t]
    \centering
        \includegraphics[width=0.32\linewidth]{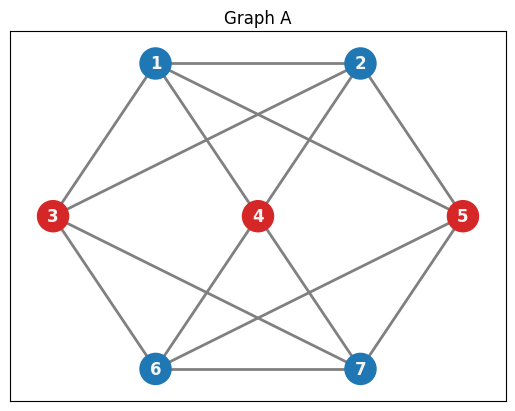}
        \includegraphics[width=0.32\linewidth]{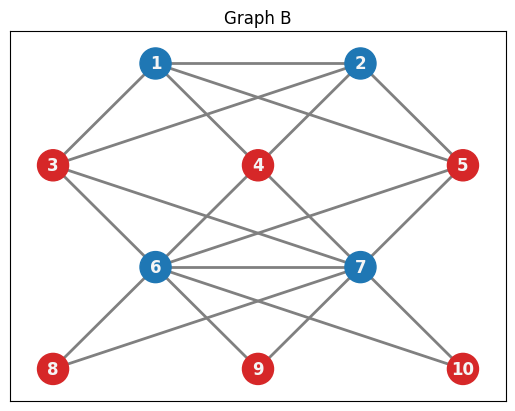}
        \includegraphics[width=0.32\linewidth]{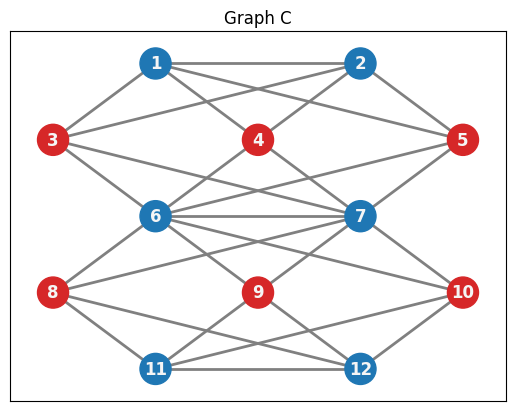}
    \caption{Visual representation of the used layered graphs.}
    \label{fig:graphs_abc}
\end{figure}

\autoref{fig:graphs_abc} shows the three layered graphs used in our experiments.
Due to the layered structure of the graph, the cost circuit shows a cascading architecture. This means that few gates have to be cut in our experiment.
In addition, determining the optimum for comparison is simple for these graphs, allowing easy verification of the obtained results.

The parameters of the QAOA algorithm are optimized classically using Simultaneous Perturbation Stochastic Approximation (SPSA)~\cite{Spall:1992}, known for its potentially faster convergence with fewer quantum circuit executions \cite{SPSAPennylane}.

\subsection{Code framework}
We used a dual-platform approach for our simulation workflow. 
PennyLane~\cite{Pennylane:2022} performs the ideal simulations, while Qiskit's Aer simulator~\cite{qiskit2024} facilitates both ideal and noisy simulations. 
The workflow first constructs circuits in Pennylane and then converts them to the OPENQASM 2.0 format using built-in tools~\cite{QASM:2017}. 
As the required mid-circuit measurements are not supported in this format, we developed a custom tool to upgrade the code to OPENQASM 3.0~\cite{QASM:2017}.
The framework then sends this OPENQASM 3.0 code to construct a Qiskit quantum circuit object.

All experiments use a fixed shot budget evenly distributed over the subcircuits. 
We measure the difference between two probability distributions $P=(p_i)_{i=1}^n$ and $Q=(q_i)_{i=1}^n$ with the Hellinger distance~\cite{HellingerDistanceChoice}:
\begin{equation}
    H(P, Q) = \frac{1}{\sqrt{2}}\sqrt{\sum_{i=1}^{n}{(\sqrt{p_i} - \sqrt{q_i}})^2}.
    \label{eq:hellinger_distance}
\end{equation}
Results are averaged of 120 and 60 independent runs for noiseless uncut and cut experiments, respectively. 
Due to significant simulation cost in noisy versions, we averaged over 20 and 10 samples for the noisy experiments, respectively. 

For the cut circuits, we assume that we have access to a quantum device of five qubits at most, requiring one, two, and three cuts for the graphs $A$, $B$ and $C$ of \autoref{fig:graphs_abc}. 
Our simulations use a noise model derived from the IBM Quantum Brisbane device~\cite{IBMQuantumExperience:2024}.

\autoref{tab:qaoa_graphs_table} shows the number of qubits used for the different graphs in both the uncut and cut quantum circuits. 
\begin{table}[h]
\centering 
\caption{Qubit requirements and the number of wirecuts for Graphs A, B, and C, demonstrating the reduction in qubits achieved through wirecutting.}
\label{tab:qaoa_graphs_table} 
\begin{tabular}{|l|c|c|c|} 
\hline
    \textbf{Graph Name} & \textbf{Wirecut Status} & \textbf{Qubits} & \textbf{Wirecut(s)} \\
    \hline
    Graph A & Uncut     & 7  & 0 \\
            & Wirecut   & 5  & 1 \\
    \hline
    Graph B & Uncut     & 10 & 0 \\
            & Wirecut   & 5  & 2 \\
    \hline
    Graph C & Uncut     & 12 & 0 \\
            & Wirecut   & 5  & 3 \\    
    \hline 
    \end{tabular}
\end{table}

\section{Results}
\label{sec:results}
This section presents the results of two distinct experiments:
In Section~\ref{sec:results_GHZ}, we investigate the efficiency and accuracy per shot of various wire-cutting techniques. 
Subsequently, in Section~\ref{sec:results_qaoa_wirecut}, we apply the randomized measurement wire-cutting technique \cite{RandomizedMeasurementsWireCutting} to QAOA for solving the MaxCut problem, evaluating its performance under both ideal and noisy simulation conditions.

\subsection{Wire cutting techniques}
\label{sec:results_GHZ}
\begin{figure}[t]
    \centering
    \includegraphics[width=0.8\linewidth]{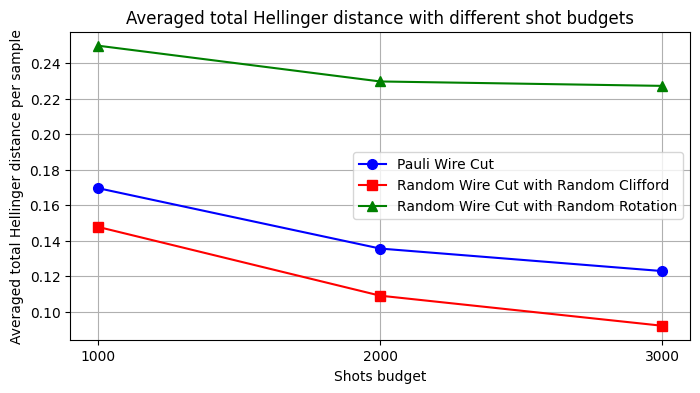}
    \caption{Total Hellinger distance averaged over 60 independent runs for three wire cutting techniques (Pauli Wire Cut, Random Wire Cut with Random Clifford, and Random Wire Cut with Random Rotation) across varying shot budgets.}
    \label{fig:results_GHZ_experiment}
\end{figure}

We tested three circuit cutting methods: an exact method based on Pauli-measurements, a randomized method that randomly chooses between a random Clifford channel and a depolarizing channel, and a method that uses random Pauli rotation gates. 
\autoref{fig:results_GHZ_experiment} shows the results for three different shot budgets. 
The randomized circuit cutting consistently outperforms the exact circuit cutting. Interestingly, applying random rotation gates before measurements and as state preparation (the third method) gives detrimental results. 

All three methods show a decreased Hellinger distance with an increased shot budget. 
As more shots reduce the statistical variance, this is as expected. 
Lower Hellinger distance implies a better performance. 
The randomized approach using Clifford gates achieves the best performance with the same shot budget, implying a lower sampling overhead. 
From the same figure, we also see that, contrary to our hopes, the random Pauli rotations have a worse performance and higher sampling overhead even than standard Pauli cutting. 
We thus conclude that the randomized approach using Clifford gates works best and therefore use that method in our second experiment. 

\subsection{Wire cutting QAOA for MaxCut}\label{results_qaoa_wirecut}
This section uses randomized circuit cutting using Clifford gates to test how well a QAOA algorithm can solve MaxCut in both a cut and uncut version and in ideal and noisy settings.
We varied the shot budget from 2000 to 4000. 

\subsubsection{Ideal Simulations}\label{sec:results_ideal_simulations}
\begin{figure}[t]
\centering
   \begin{subfigure}{\textwidth}
   \centering
   \includegraphics[width=\linewidth]{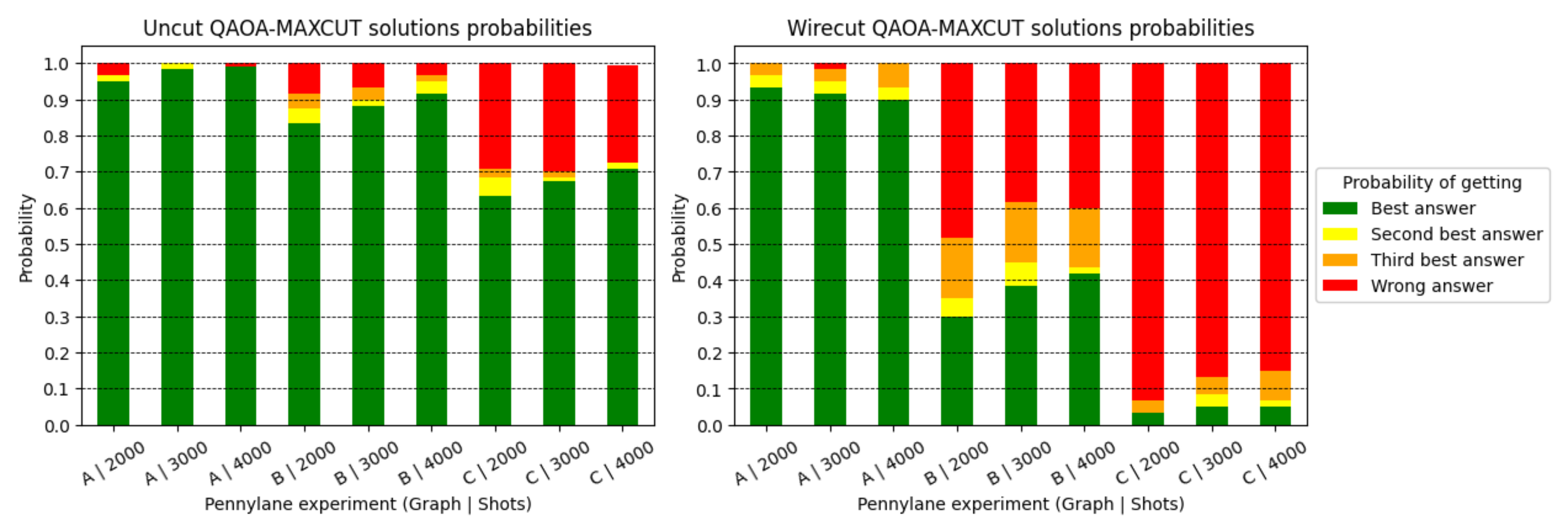}
   \caption{Results for the QAOA-MaxCut experiment with ideal simulation.}
   \label{fig:pennylane_ideal_uncut_and_wirecut} 
\end{subfigure}
\begin{subfigure}{\textwidth}
    \centering
    \includegraphics[width=\linewidth]{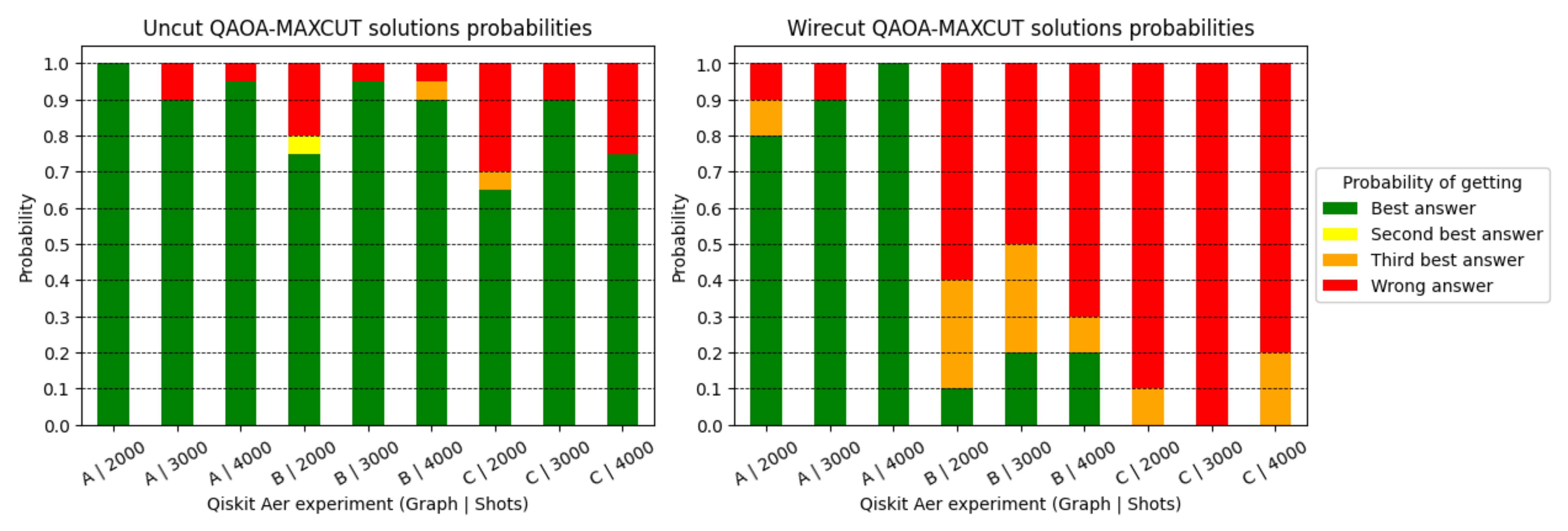}
    \caption{Results for the QAOA-MaxCut experiment with noisy simulation. For the noise model of the IBM Brisbane quantum device was used.}
    \label{fig:qiskit_noisy_uncut_and_wirecut}
\end{subfigure}
\caption{Numerical results on the performance of circuit cutting for QAOA. The left figures show results using uncut circuits, and the right figures show results using circuit cutting. Each bar shows the fraction of best, second best, third best, and wrong answer measured three graph instances and different shot budgets.}
\end{figure}

\autoref{fig:pennylane_ideal_uncut_and_wirecut} shows how often the optimal cut is found in both an uncut and a cut run of the QAOA algorithm in an ideal simulation. 
We considered the three graphs shown in \autoref{fig:graphs_abc}.

Again, as shot budget increases, the optimal answer is found more often. 
As the complexity of the graphs increases, the success probability does however decrease. 
For the wire cut circuits, we see that especially for the larger graphs, suboptimal answers are found most of the time. 
This is quickly explained by the number of wire cuts applied (two for graph $B$ and three for graph $C$), while fixing the shot budget. 
As a result, statistical variances have a greater effect on the overall performance. 
For improved performance, the shot budget should be significantly increased for the larger graphs: 
Every subcircuit requires sufficient samples to obtain the correct outcomes with high probability. 

\subsubsection{Noisy Simulations}\label{sec:results_qaoa_wirecut}
This section details the results of our noisy simulations for QAOA-MaxCut, comparing the performance of uncut and wirecut quantum circuits for the three graphs shown in \autoref{fig:graphs_abc}. 

\autoref{fig:qiskit_noisy_uncut_and_wirecut} shows the results of the experiments with noisy simulations. We see that noisy simulations for the wirecut circuits are significantly worse than the uncut version. 
Contrary to what we expect, increasing the shot budget did not always result in higher accuracy. 
We believe this unexpected behavior stems from the fixed shot budget. 
With an increasing number of circuit cuts, the number of simulated subcircuits also increases. 
Each of these subcircuits gets assigned a smaller shot budget, making them more susceptible for statistical inaccuracies. 
The fact that for Graph $C$, the correct outcome was never found supports this feeling. 

\section{Conclusion}
\label{sec:conclusion}
In this work we considered circuit cutting, a technique to break down a quantum circuit in smaller blocks. 
In all of our experiments, we fixed the shot budget, regardless of whether a circuit cut was used or not. 
In theory, an exponential number of shots is needed in the number of cuts performed.
As a result, we expect statistical variance in our results, especially when multiple shots are being used. 

We considered three different circuit cutting techniques and found that circuit cutting based on randomized Clifford gates achieves the best performance.
We tested this using a circuit to prepare a GHZ state, where additionally random Pauli rotation gates are applied in between the CNOT gates. 

In the second experiment we considered the effect of QAOA for solving a MaxCut problem. 
Here, we considered three different graphs, which required a different number of circuit cuts. 
Especially for the larger graphs, those that require two or even three circuit cuts, the performance lacks significantly compared to the uncut version. 
This effect is magnified in the noisy simulations. 
As mentioned before, we expect that this results in part from the need of an exponential number of shots, whereas a fixed number of shots was used.
Overall, it can be concluded that applying wire cutting to QAOA reduces its accuracy significantly, though it enables execution on quantum hardware with fewer qubits.
Furthermore, this loss in accuracy can be mitigated by increasing the shot budget for larger graphs with multiple wire cuts in ideal simulations but remains unknown in noisy circumstances.

One can imagine that the impact of individual terms in Equation~\eqref{eq:pauli_wire_cut} differs. 
Some terms have a large impact, while others have a smaller impact. 
In practice, it might therefore be advantageous to use a skewed shot budget, with more shots being allocated to large impact terms. 
This is an interesting topic for follow-up research. 

For future research, we believe that larger sample sizes can help reduce statistical variances, this includes both larger shot budgets and averaging results over multiple independent runs. 
Especially the noisy simulations were computationally heavy, particularly due to the need for mid-circuit measurements.
Next, it is vital to test the methods on actual hardware to see the performance in practice. 
Future research can also incorporate error-mitigation techniques, such as Pauli-twirling~\cite{Wallman:2016} or zero-noise extrapolation~\cite{TemmeBravyiGambetta:2017,Li:2017}, to reduce the effect of noise.




\section*{Acknowledgments}
This work is part of the project \textit{Divide and Quantum} `D\&Q' NWA.1389.20.241 of the program `NWA-ORC', which is partly funded by the Dutch Research Council (NWO). 

\bibliographystyle{splncs04}
\bibliography{QUESTIS-221}
    
\end{document}